\documentclass[12pt]{iopart}
\usepackage{a4}
\usepackage{graphicx}
\usepackage{indentfirst}
\usepackage{pifont}
\usepackage{pslatex}
\begin{document}
\title{A First Comparison Between LIGO and Virgo Inspiral Search Pipelines}
\author{L.~Blackburn$^3$, 
F.~Beauville$^5$, 
M.-A.~Bizouard$^7$, 
L.~Bosi$^8$, 
P.~Brady$^4$, 
L.~Brocco$^9$, 
D.~Brown$^{2,4}$,
D.~Buskulic$^5$, 
S.~Chatterji$^2$, 
N.~Christensen$^1$, 
A.-C.~Clapson$^7$, 
S.~Fairhurst$^4$, 
D.~Grosjean$^5$, 
G.~Guidi$^6$, 
P.~Hello$^7$, 
E.Katsavounidis$^3$, 
M.~Knight$^1$, 
A.~Lazzarini$^2$, 
F.~Marion$^5$, 
B.~Mours$^5$, 
F.~Ricci$^9$, 
A.~Vicer\'e$^6$, 
M.Zanolin$^3$ \\[.3cm]
\begin{center}The joint LIGO/Virgo working group\end{center}}
\vspace{.3cm}
\address{$^1$ Carleton College, Northfield MN 55057 USA}
\address{$^2$ LIGO-California Institute of Technology, Pasadena CA 91125 USA}
\address{$^3$ LIGO-Massachusetts Institute of Technology, Cambridge, Massachusetts 02139 USA}
\address{$^4$ University of Wisconsin - Milwaukee, Milwaukee WI 53201 USA}
\address{$^5$ Laboratoire d'Annecy-le-Vieux de physique des particules, 
Chemin de Bellevue, BP 110, 74941 Annecy-le-Vieux Cedex France }
\address{$^6$ INFN - Sezione Firenze/Urbino
Via G.Sansone 1, I-50019 Sesto Fiorentino; and/or Universit\`a di
Firenze, Largo E.Fermi 2, I-50125 Firenze and/or Universit\`a di Urbino, Via S.Chiara 27, I-61029 Urbino Italia}
\address{$^7$ Laboratoire de l'Acc\'el\'erateur Lin\'eaire
(LAL), IN2P3/CNRS-Universit\'e de Paris-Sud, B.P. 34, 91898 Orsay Cedex France}
\address{$^8$ INFN Sezione di Perugia and/or  Universit\`a di Perugia,
Via A. Pascoli, I-06123 Perugia Italia}
\address{$^9$ INFN, Sezione di Roma  and/or Universit\`a ``La Sapienza",  P.le A. Moro 2, I-00185, Roma Italia}
\begin{abstract}
This article reports on a project that is the first step the LIGO Scientific 
Collaboration and the Virgo Collaboration have taken to prepare for the 
mutual search for inspiral signals. The project 
involved comparing the analysis pipelines of the two collaborations on 
data sets prepared by both sides, containing simulated noise and injected 
events. The ability of the pipelines to detect the injected events was 
checked, and a first comparison of how the parameters of the events were 
recovered has been completed.
\end{abstract}
\maketitle
\section*{Introduction}
The LIGO Scientific Collaboration and the Virgo Experiment have agreed to pursue 
a joint search for binary inspiral signals~\cite{proposal}. The proposal 
defines goals and steps to scientifically analyze data in the search 
for burst events and for coalescence and mergers of compact binary systems. 
To achieve these goals, both collaborations have undertaken to work together 
on a series of preparatory data analysis projects of increasing complexity. 
\par
This article reports on the work done on the first test project in
the search for inspiral signals from compact binary 
systems. The primary purpose of the first project was to gain a 
better understanding of the data analysis algorithms and procedures employed 
by each group, and to develop a common analysis language. 
\par
The first project involved learning how to exchange interferometer data and 
trigger files, and gaining experience in analyzing each others' data. It 
also involved comparing the various search algorithms and their 
implementations for detection efficiency, as well as computational efficiency. 
In particular, we wanted to verify that both groups' detection algorithms 
identify the same injected events in the data streams and that the recovered 
parameters of these events are in good agreement with the injection 
parameters, in order to establish confidence in both our injection and 
detection procedures. This allowed us to compare alternative implementations 
of matched filtering for a binary inspiral search. 
A similar project has been carried 
out simultaneously for burst events~\cite{burst_paper}.

\par
Each collaboration created three hours of single interferometer simulated data
matching their design sensitivity. Additionally, each group provided a series 
of inspiral injections to be added to the 
data. Both the LSC and Virgo analysis codes were then run over both data sets, 
over the same source mass range and with the same threshold on the signal to 
noise ratio. Triggers from the searches have been exchanged, to compare 
detection efficiency, computational cost and parameter recovery. 
\par
The first section of this article presents the features of the simulated 
data that have been generated by each collaboration for the project. 
The analysis pipelines used to search the data are described in section 2.
Section 3 is devoted to the trigger production, and section 4 presents a 
detailed comparison of the events identified by the different analysis codes. 
\section{Simulated Data}
\begin{figure}[h!] 
\begin{center}
\includegraphics[height=5.cm]{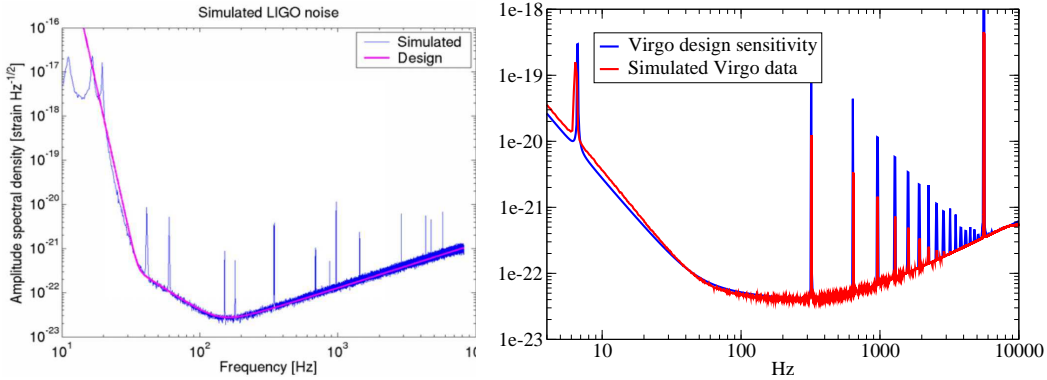}
\includegraphics[height=5.cm]{sensVirgo.eps}
\end{center}
\caption{Amplitude spectral density of simulated noise matching the 
design sensitivity of LIGO (left) and Virgo (right), in strain $/ \sqrt{Hz}$.}
\label{sensitivity}
\end{figure}

To serve as a benchmark for this first project, each collaboration produced 
three hours of single interferometer data consisting of colored stationary 
Gaussian noise at their own sample rate (16384 Hz for LIGO and 20000 Hz 
for Virgo) with a spectrum matching their design sensitivity, including 
expected narrowband features. Figure~\ref{sensitivity} shows the amplitude 
spectral density of each set of simulated data.
\par
In addition to the detector noise, each group provided a series of 
non-coincident, optimally oriented, neutron star inspiral waveforms to be 
injected in the data. The events were generated using second order post 
Newtonian waveforms. 
\par
A number of 26 such events were randomly injected in the LIGO data, 
with source masses either $[1.4 M_{\odot},1.4 M_{\odot}]$ or 
$[1 M_{\odot},2 M_{\odot}]$, with $M_{\odot}$ the mass of the Sun. 
The source was assumed to be located at a distance either 20, 25, 30 or 
35~Mpc from the detector. The waveforms were generated with a starting frequency 
of 40~Hz, consistent with the interferometer sensitivity, and injected 
with an average period of about 400~seconds. 
\par
In the Virgo data, 11 inspiral events with source masses 
$[1.4 M_{\odot},1.4 M_{\odot}]$ were randomly injected with an average 
period of about 900~seconds. The starting frequency of the waveform was 24~Hz, 
and the events were injected with a signal to noise ration (SNR) of 10, 
corresponding to a distance of 24.8~Mpc between the source and the detector.
\par
All data were stored in frame format, and only calibrated strain data 
were exchanged. Details of the injected signals were also exchanged in 
advance. 
\section{Analysis Pipelines}
Each group analyzed both sets of LIGO and Virgo simulated data, using their 
own analysis pipeline. In this section we give a short description of the 
analysis codes used. They are all based on matched filtering implemented 
on a bank of templates covering some given space for the source masses.
\subsection{LSC Pipeline}
A detailed description of the LSC inspiral search pipeline can be found 
in~\cite{S1_inspiral}. We only summarize here the main features of the 
algorithm. 
\par
The data set is split into analysis chunks of 15 overlapping segments. 
The length of each segment is chosen so as to be at least four times the 
duration of the longest template in the bank, which depends on the starting 
frequency of the search. 
For each analysis chunk, a bank of templates is created, generating templates 
in the frequency domain using second order post Newtonian waveforms. 
\par
The data are filtered with each template in the bank, and a trigger is 
recorded each time the maximum of the SNR of a filter output exceeds 
some threshold. Maxima above threshold separated in time by less than the 
length of the template are considered the same trigger, but triggers from 
different templates are recorded individually. 
\subsection{Virgo Pipeline}
For this project, the multi-band search (MBTA)~\cite{Moriond}, which is one 
of the template-based analyzes implemented by the Virgo Collaboration, has 
been used. In this approach, the 
templates are split for efficiency into low and high frequency parts and 
then combined together, in a hierarchical way.
\par
For each frequency band, a bank of ``real'' templates is created, generating 
templates first in the time domain using second order post Newtonian 
waveforms. These templates are called real because they are actually used to 
filter the data. The analysis is done on data segments long enough to 
accommodate at least twice the longest template length.
\par
On the full frequency band, a bank of ``virtual'' templates is created with 
the same criteria, but the templates are not used directly in matched filters. 
For each virtual template, the output of the corresponding filter is built 
coherently from the outputs of the filters based on the real templates 
associated to the virtual template in each frequency band.  
\par
A trigger is recorded each time the maximum of the SNR of a (virtual) filter 
output exceeds some threshold. The triggers are then clustered both in time 
and over the template bank; triggers issued by different templates but with 
matching ending time are considered the same event. 
\par
A "flat search" more similar to the LSC pipeline has also been
implemented in Virgo, and was run on both data sets. The triggers 
obtained are consistent with those resulting by the MBTA analysis
and for the sake of article's clarity we do not report further details.
Readers interested in the "flat search" approach can refer to the paper
by L.Bosi et al. in these proceedings~\cite{bosi}.
\section{Production of Trigger Lists}
The LSC pipeline and the MBTA pipeline were used to analyze both data sets, 
with common search parameters agreed upon beforehand by the two groups. The 
parameters of the analysis are summarized in table~\ref{prod_parameters}. 
The source mass space explored was 1-3 M$_{\odot}$. The template banks were 
generated matching a grid in the mass space created with a minimal match 
criterion of 95\%, ensuring that no event in that mass space should be 
detected with a SNR loss greater than 5\%.
\par
The starting frequency $f_{low}$ used to build the templates was set to different 
values to analyze the LIGO data and the Virgo data, to be consistent 
with the SNR accumulation driven by the sensitivity curves of both 
experiments; $f_{low}$ was set to 40~Hz for LIGO data and to 30~Hz for 
Virgo data. In addition, the MBTA code was run with a splitting frequency 
between the low and high frequency bands chosen so as to share in an 
approximately equal way the SNR between the two bands (see 
table~\ref{production}). Triggers were recorded when the SNR exceeded a 
threshold of 6.
\begin{table}[h!]
\begin{center}
\begin{tabular}{|l|c|c|}
\hline
 & LIGO data & Virgo data \\ \hline
 Mass range & 1-3 M$_{\odot}$  & 1-3 M$_{\odot}$ \\ \hline
 Grid minimal match & 95\% & 95\%  \\ \hline
 Starting frequency $f_{low}$ & 40 Hz & 30 Hz \\ \hline
Longest template duration & $\sim$ 45 seconds & $\sim$ 96 seconds \\ \hline
 SNR threshold & 6 & 6 \\ \hline
\end{tabular}
\end{center}
\caption{Common search parameters for the LSC and MBTA pipelines.}
\label{prod_parameters}
\end{table}
\par
The layout of the template banks depends on the noise power spectral density 
of the instrument, and on the value chosen for $f_{low}$. The way the LSC 
designs the grid used to place the templates is described 
in~\cite{Owen_Sathya}. The Virgo group creates the grid according to a 
2D contour reconstruction technique based on the parameter space 
metric~\cite{Damir}. The two methods lead to numbers of templates that 
differ by about 40\%, as is reported in table~\ref{production}.
\begin{table}[h!]
\begin{center}
\hspace*{-1cm}
\begin{tabular}{|l|c|c|c|c|}
\hline
& LSC pipeline & MBTA pipeline & LSC pipeline & MBTA pipeline \\ 
&  on LIGO data & on LIGO data & on Virgo data & on Virgo data\\ \hline
Number of templates & $\sim$ 3500 & $\sim$ 1900 & $\sim$ 9500 & $\sim$ 6100 \\ \hline
Band splitting frequency & not relevant & 130 Hz & not relevant & 95.3 Hz \\ \hline
Number of jobs & 6 & 1 & 3 & 7 \\ \hline
Type of processor & 1 GHz Pentium II & Xeon 2 GHz & Xeon 2.66 GHz & 2.4 GHz Pentium IV \\ \hline
Total memory & & 0.9 GByte & & 4.5 GBytes \\ \hline
Total processing time & $\sim$ 46 hours & $\sim$ 13 hours 
& $\sim$ 88 hours & $\sim$ 28 hours \\ \hline
$\frac{{\rm processing\ time} \times{\rm processor\ speed}}{{\rm number\ of\ templates}}$
& \raisebox{0cm}[.6cm][.4cm]{$\sim$ 47 s\ GHz} & $\sim$ 49 s\ GHz & $\sim$ 89 s\ GHz & $\sim$ 40 s\ GHz \\ \hline
\end{tabular}
\end{center}
\caption{Configuration and computing cost of each analysis.}
\label{production}
\end{table}
\par
Table~\ref{production} also provides information about the way the 
production was done for the two pipelines and each data set: on how many jobs 
the production was split (the LSC pipeline analyzes different time periods 
in different jobs, whereas the MBTA explores different regions of the 
parameter space), which type of processor was used and how much resources 
were needed (memory, total processing time after initialization).
\par
In order to compare the performances of the two pipelines, the table also 
quotes the amount of time required to analyze one template, normalized by the 
speed of the processor. The speeds of all the analyses are about the same, 
apart from the LSC pipeline running on Virgo data, presumably because it did 
not use a number of points which was a power of 2, and this would have slowed
down the fast Fourier transforms. Regarding the MBTA pipeline, the version 
of the code used was optimized for memory.
\section{Comparison of Triggers}
Since both groups use different formats to record their triggers, some 
specific software was developed to convert triggers from one format to the 
other. 
\subsection{Injection Identifications}
The first point was to compare how both groups' detection procedures 
were able to identify the inspiral injections present in the data streams. 
In each trigger list, an element is tagged {\em true} if the ending time 
of the event matches the ending time of the injected event within $\pm$~20~ms.
The detection efficiency and the identification overlap of the two pipelines 
are very good: The 11 events injected in the Virgo data are detected by both 
pipelines. In the LIGO data stream, there is one event out of 26 which is 
missed by both analyzes, and one event which is identified by the LSC pipeline 
but missed by the Virgo pipeline. This is interpreted as a threshold effect
since both events were injected at a distance of 35~Mpc, leading to 
an expected low value of 6.40 for the SNR. 
\par
An issue was to associate the triggers produced by the LSC pipeline
with those produced by MBTA. The LSC triggers are not clustered over 
the template bank, so that many of them usually 
correspond to a given injection. On the other hand, MBTA 
usually produces a single trigger per injection due to the clustering over 
templates. This becomes important to further compare the detection parameters 
of events identified by the codes of the two groups. For each injection, 
the trigger with highest SNR among the LSC associated triggers was kept to 
be compared to the corresponding MBTA trigger. 
\subsection{Signal to Noise Ratio}
\begin{figure}[h!] 
\begin{center}
\includegraphics[height=6cm]{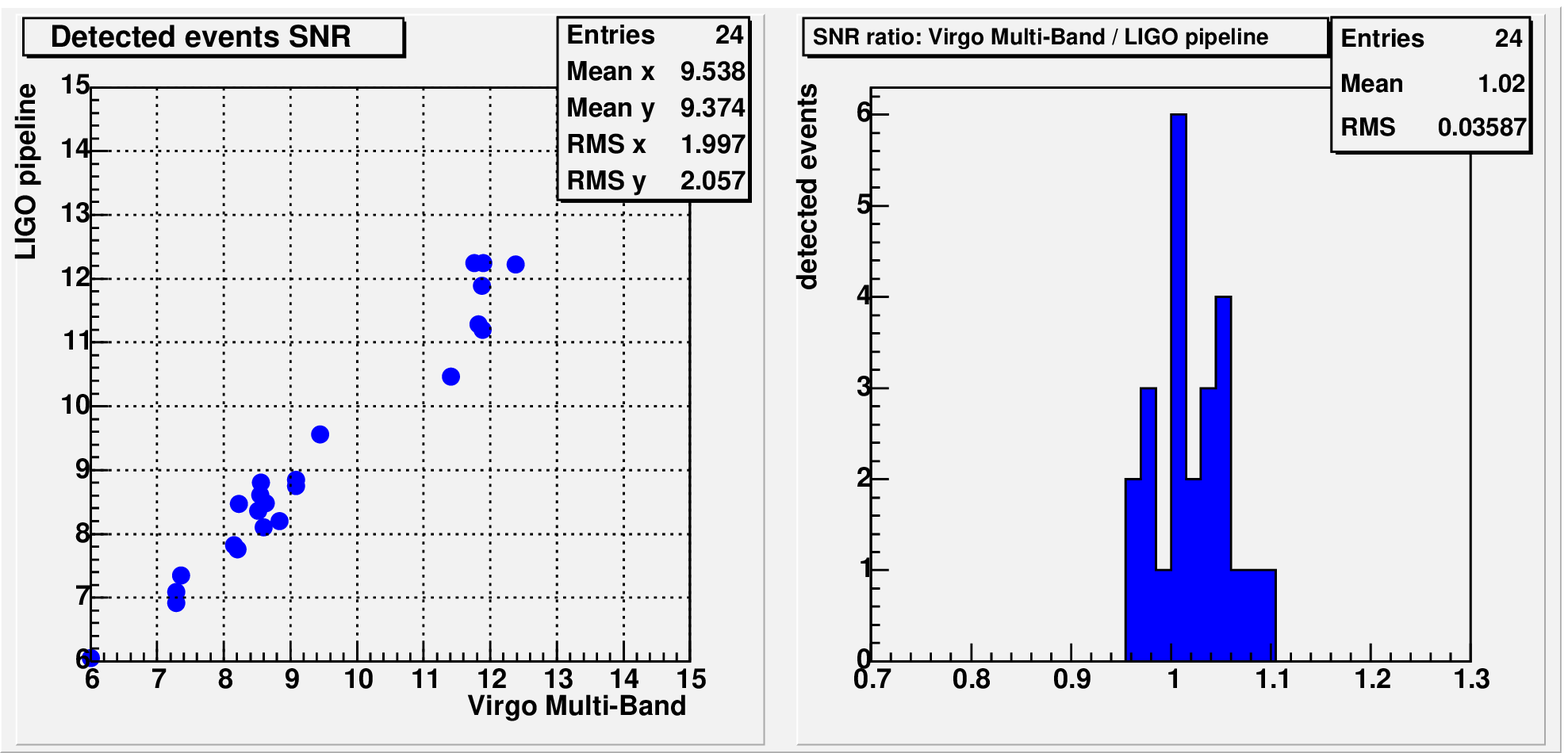}
\includegraphics[height=6cm]{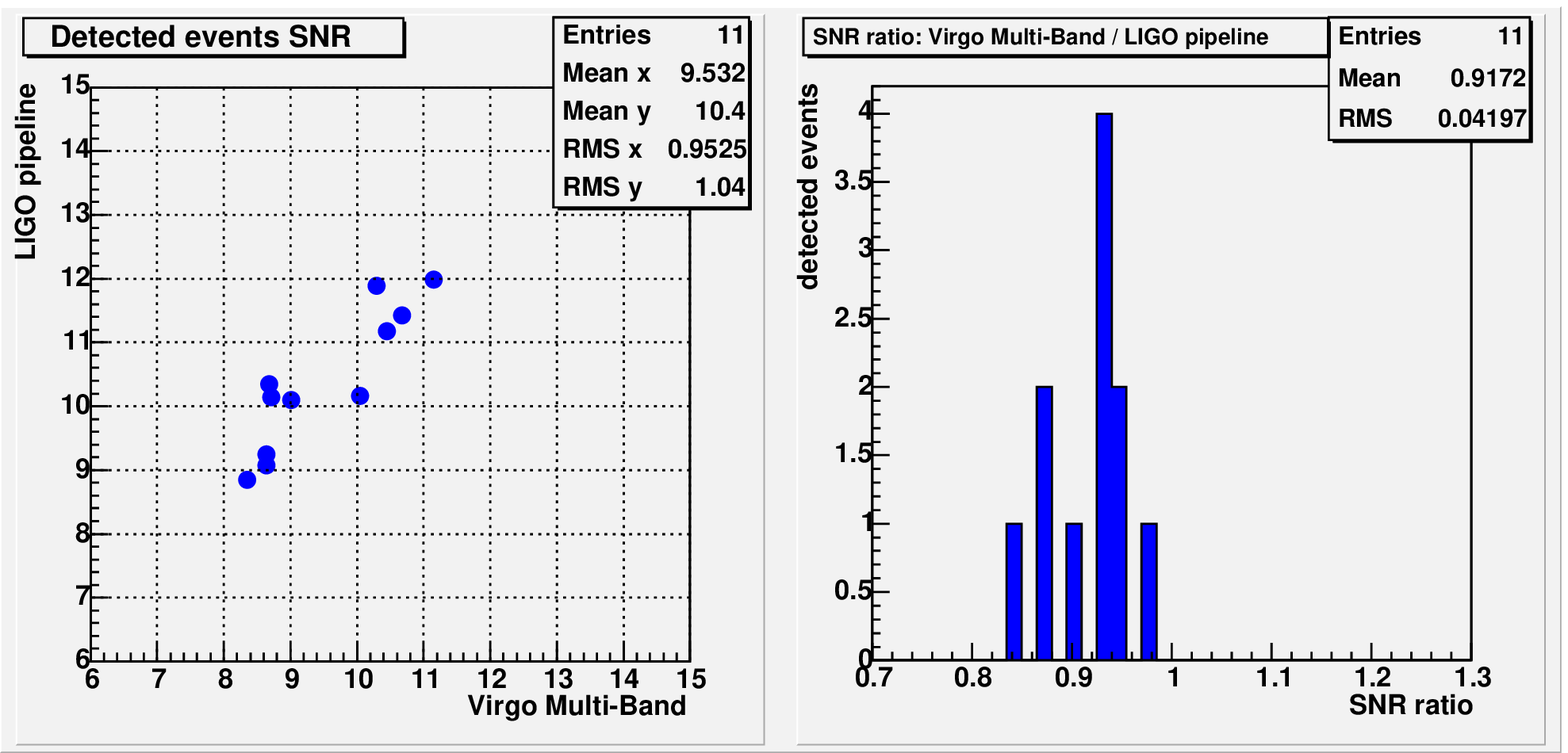}
\vspace*{-.2cm}
\end{center}
\caption{Comparison of the measured signal to noise ratio. The top row corresponds to the LIGO produced data, while the bottom row is Virgo produced.}
\label{snr}
\vspace*{-.2cm}
\end{figure}

For each injected event coincidentally detected by both pipelines, the 
measured signal to noise ratio was compared. The result of the comparison 
is shown in figure~\ref{snr}. On both data sets, there is a good correlation 
and general agreement between the SNRs measured by the LSC and MBTA pipelines.
\par
In the case of the Virgo data, the two pipelines appear to be slightly biased 
in opposite directions, so that there is an average 8\% discrepancy between 
the measured SNRs. 
The statistically significant part of the difference can be attributed to
differences in the way the template grids are generated. In particular,
while the LSC grid has a point very close to $[1.4 M_{\odot},1.4 M_{\odot}]$, 
the grid used by MBTA does not have a template very close to this point where
signals were injected. The closest template has a match only 
slightly above the 95\% required minimal match. 
\subsection{Distance}
\begin{figure}[h!] 
\begin{center}
\includegraphics[height=6cm]{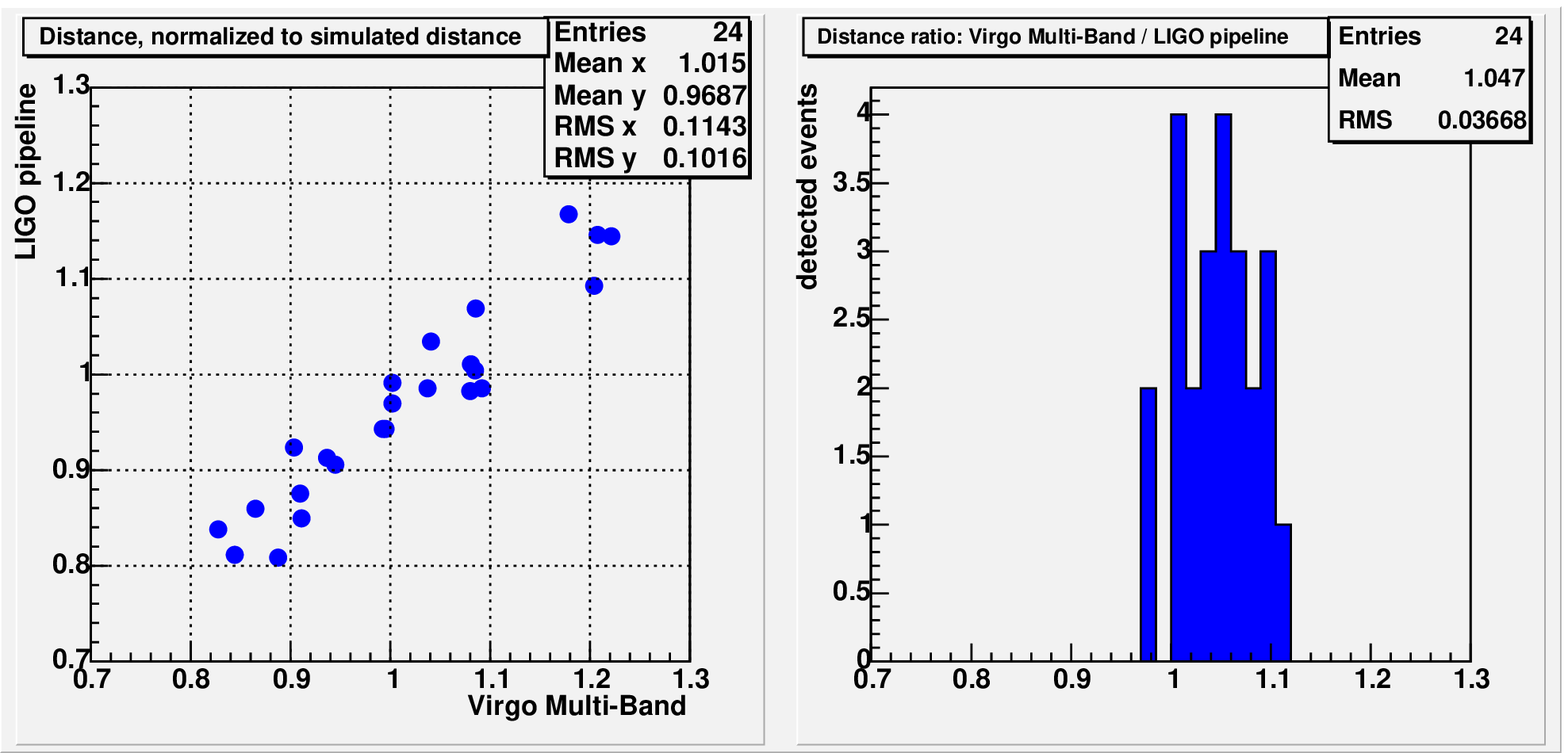}
\includegraphics[height=6cm]{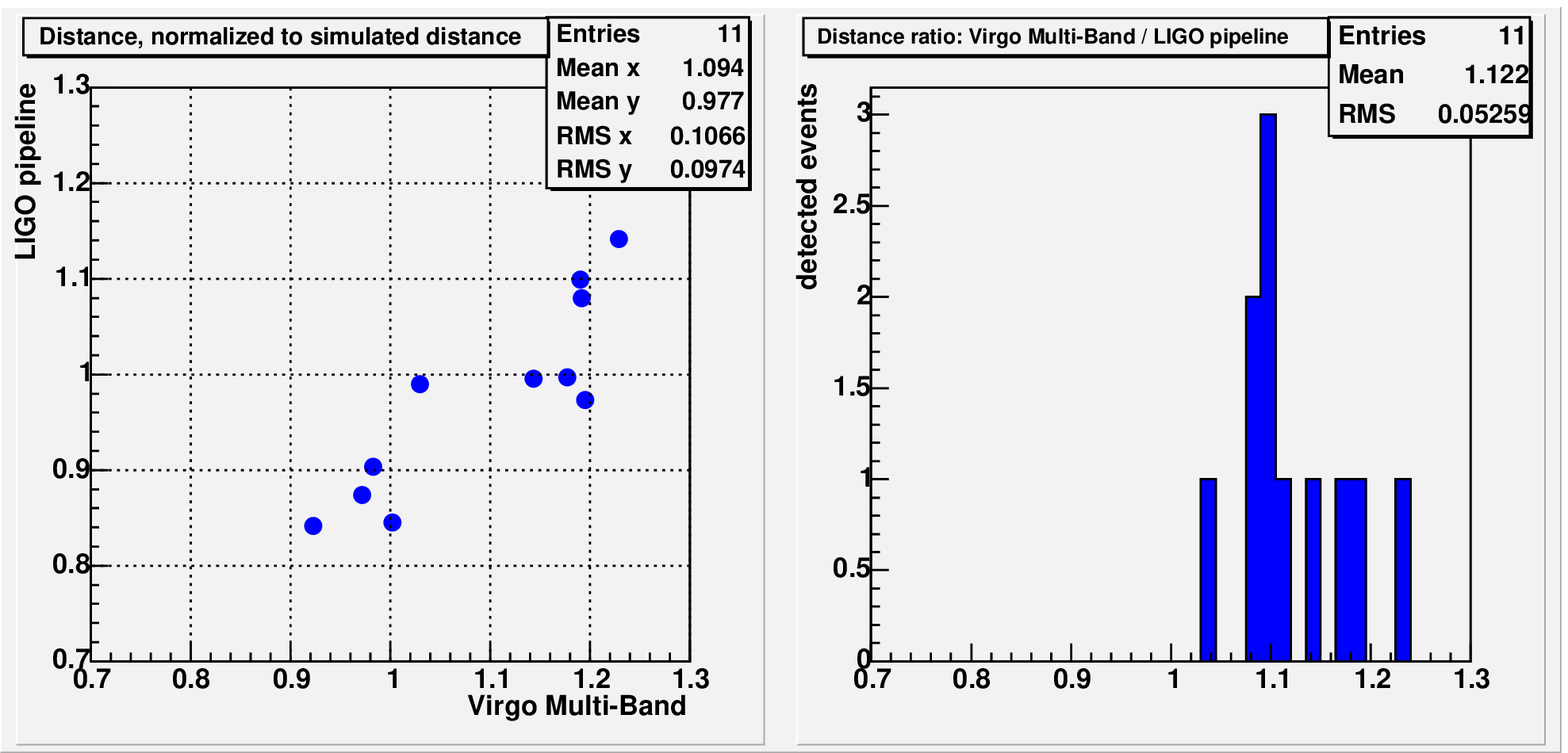}
\end{center}
\caption{Comparison of the measured source distance, normalized to the 
simulated distance. The top row corresponds to the LIGO produced data, while 
the bottom row is Virgo produced.}
\label{distance}
\end{figure}

The distances to the source measured by both pipelines were also compared, 
as shown in figure~\ref{distance}. As expected, the same correlation and 
general agreement as in the SNR case is observed. The quality of the agreement 
is somewhat degraded, however, and in the worst case of the Virgo data, is 
at the 12\% level. 
\par
The systematic effects at the origin of this bias in the distance recovery 
are yet to be investigated. In the MBTA case, for instance, it is clear that 
the distance overestimation exceeds the loss in the measured SNR, which is 
not consistent. The way triggers from the two pipelines are associated for 
comparison could also have an impact. 
\subsection{Arrival Time}
\begin{figure}[h!] 
\begin{center}
\includegraphics[height=6cm]{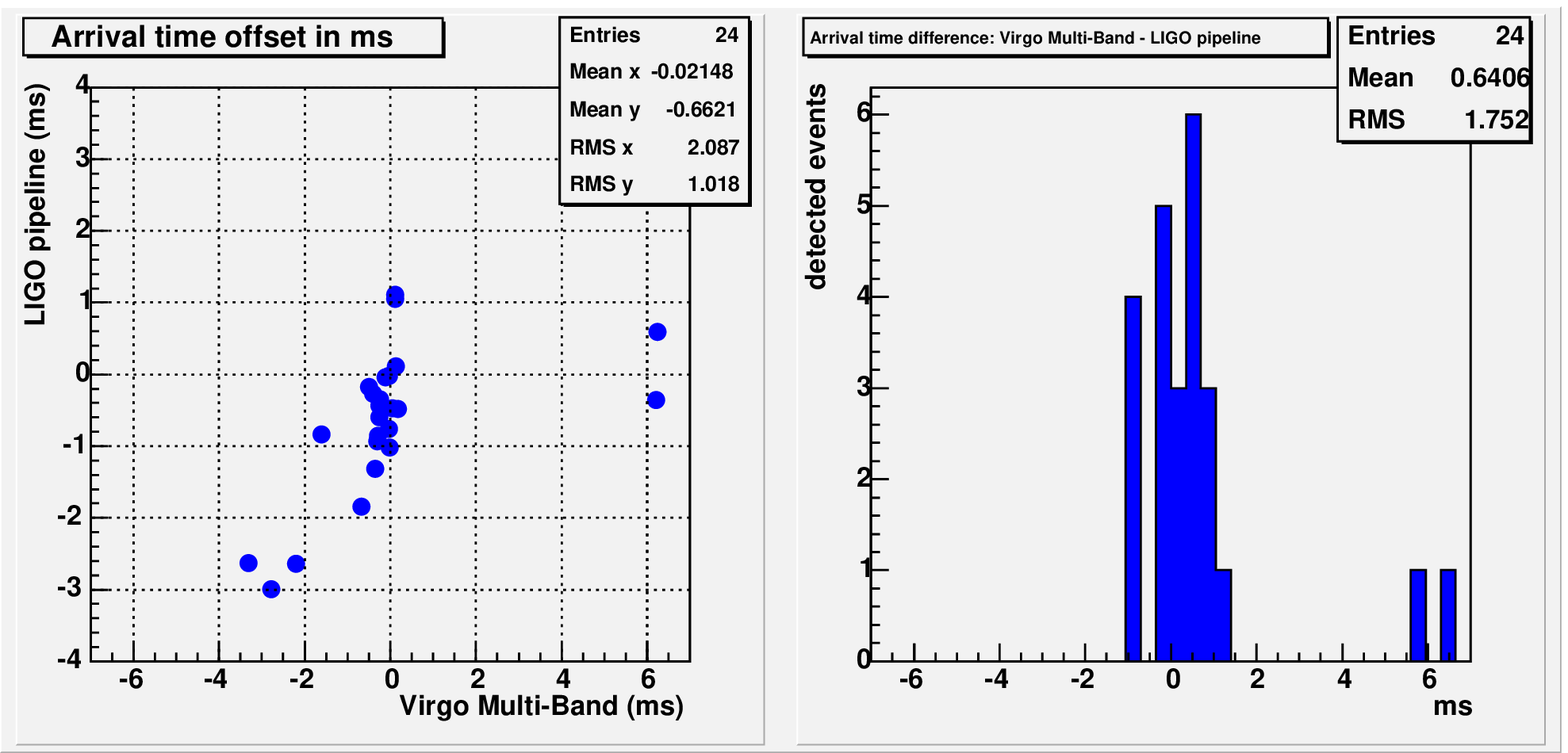}
\includegraphics[height=6cm]{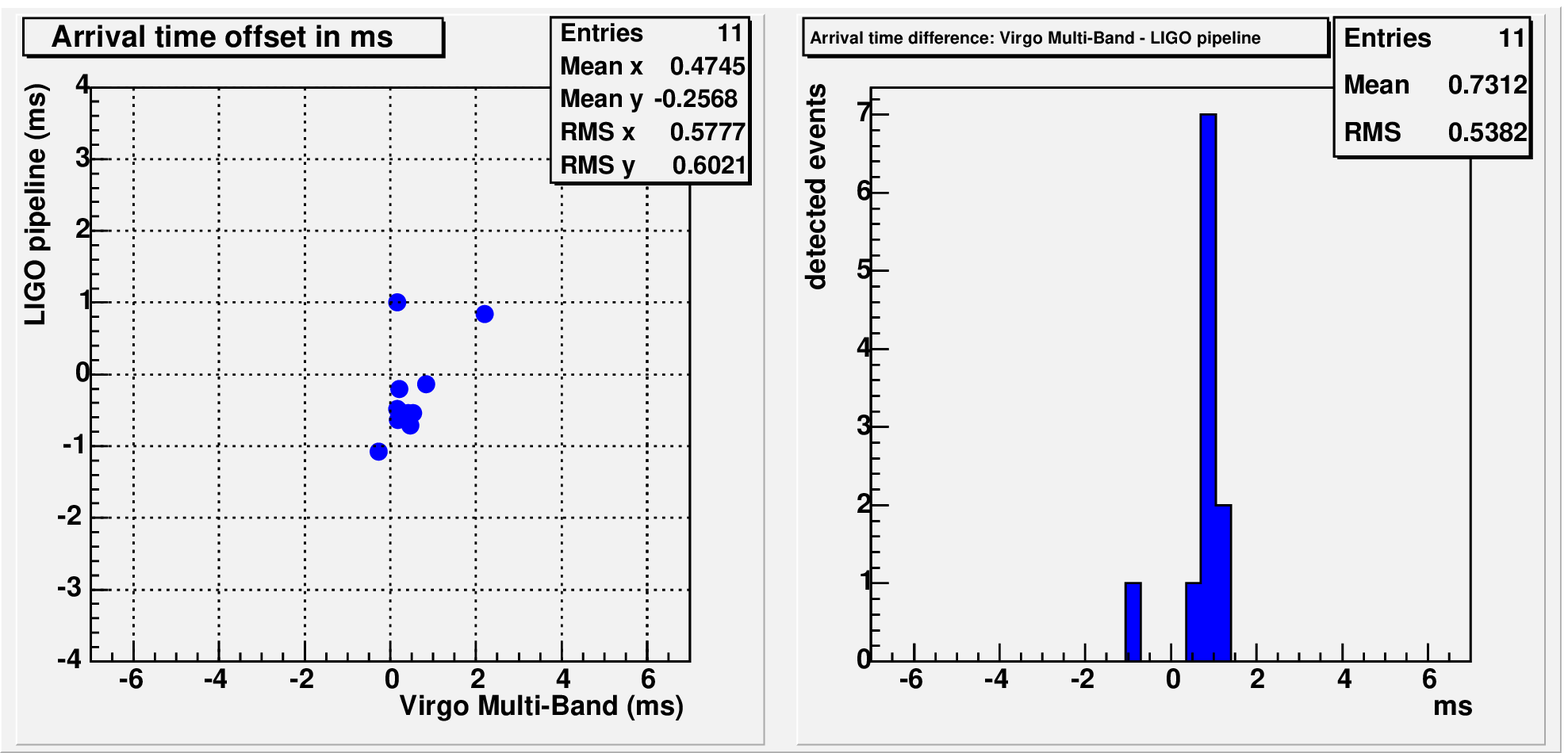}
\end{center}
\caption{Comparison of the measured arrival time. The top row corresponds to the LIGO produced data, while the bottom row is Virgo produced.}
\label{time}
\end{figure}

A crucial parameter in view of coincident analysis is the arrival time of 
the detected events. It is essential that the two pipelines agree on this 
parameter. Figure~\ref{time} shows a comparison of the arrival times quoted 
by the LSC and MBTA pipelines.
\par
Interestingly, the plots exhibit some correlation also for the arrival time. 
In the case of LIGO data, the MBTA pipeline is clearly off for two events 
detected by the same template which does not behave well as far as the 
arrival time is concerned. Apart from these two events, which are detected 
with a low SNR, the agreement between 
the measured arrival times is at the 1~ms level. 
\subsection{Source Mass}
\begin{figure}[h!] 
\begin{center}
\includegraphics[height=5.5cm]{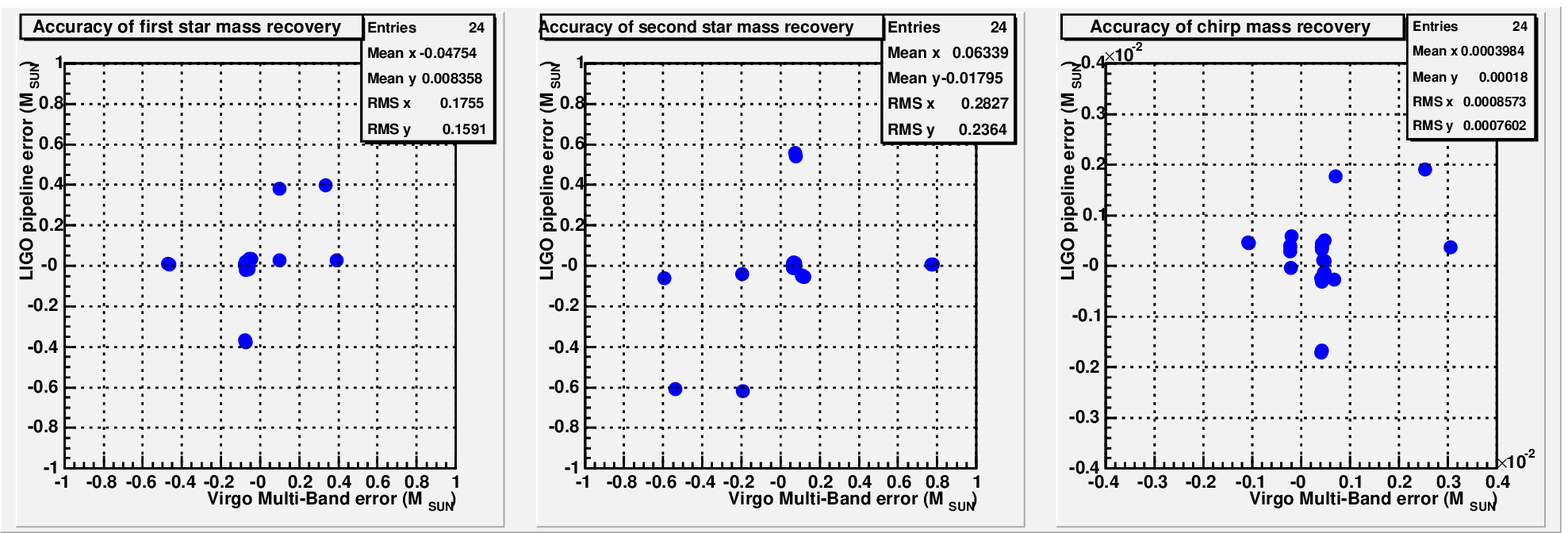}
\includegraphics[height=5.5cm]{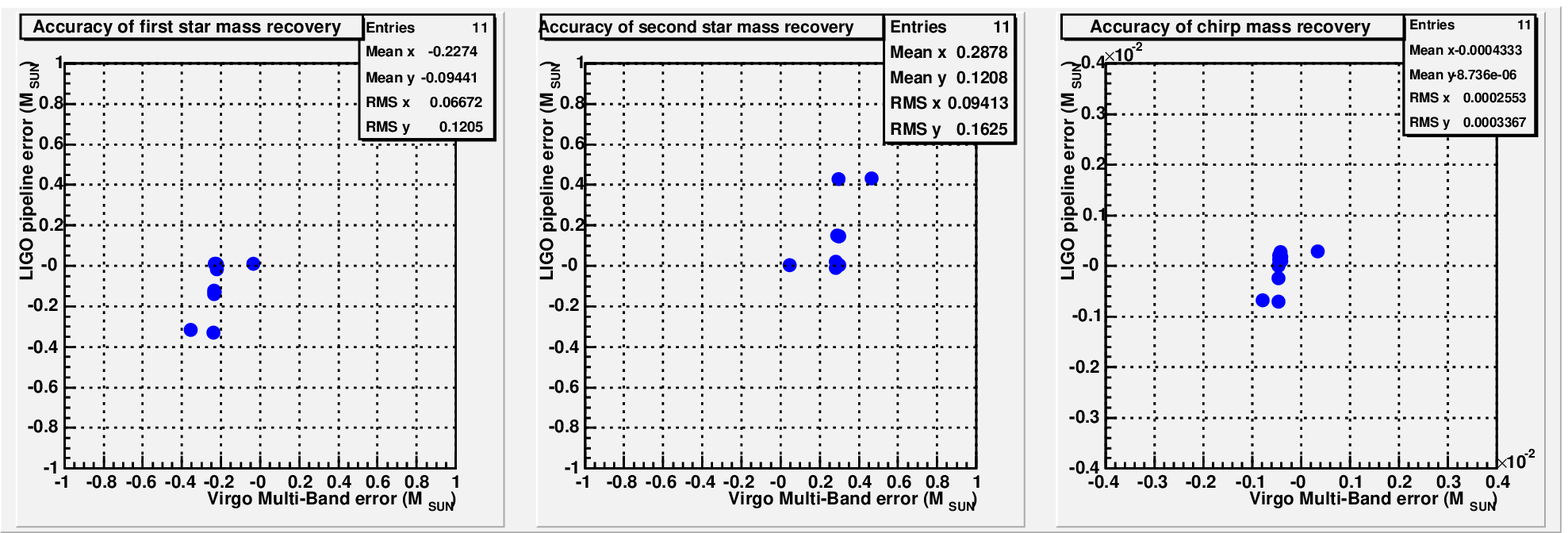}
\end{center}
\caption{Comparison of the measured source mass parameters. The top row corresponds to the LIGO produced data, while the bottom row is Virgo produced.}
\label{masses}
\end{figure}

The last test was to see how accurately the mass parameters of an injected 
inspiral signal could be reconstructed. Figure~\ref{masses} shows that 
neither pipeline is able to obtain the component masses particularly 
accurately. On the other hand, a reliable estimate of the chirp mass,
$m_{chirp}=(m_1 m_2)^{3/5} / (m_1 + m_2)^{1/5}$, can 
be obtained. For the two data sets, we see that the accuracies of both 
pipelines are comparable. 
Furthermore we recover the chirp mass very well~-~errors less than 
$4 \cdot 10^{-3}$ in LIGO data and $1 \cdot 10^{-3}$ in Virgo data (in contrast
to previous sections, the two numbers we quote here are based on the different 
data sets rather than the different pipelines, which actually give
comparable results). Thus, 
the chirp mass seems a natural parameter to use in coincidence tests. The 
better accuracy in Virgo data is due to better low frequency sensitivity and 
hence longer waveforms. The errors in the chirp 
mass appear uncorrelated between the pipelines, and are probably due to the 
specific template placement algorithms used.
\par
\begin{figure}[h!] 
\begin{center}
\includegraphics[height=5.cm]{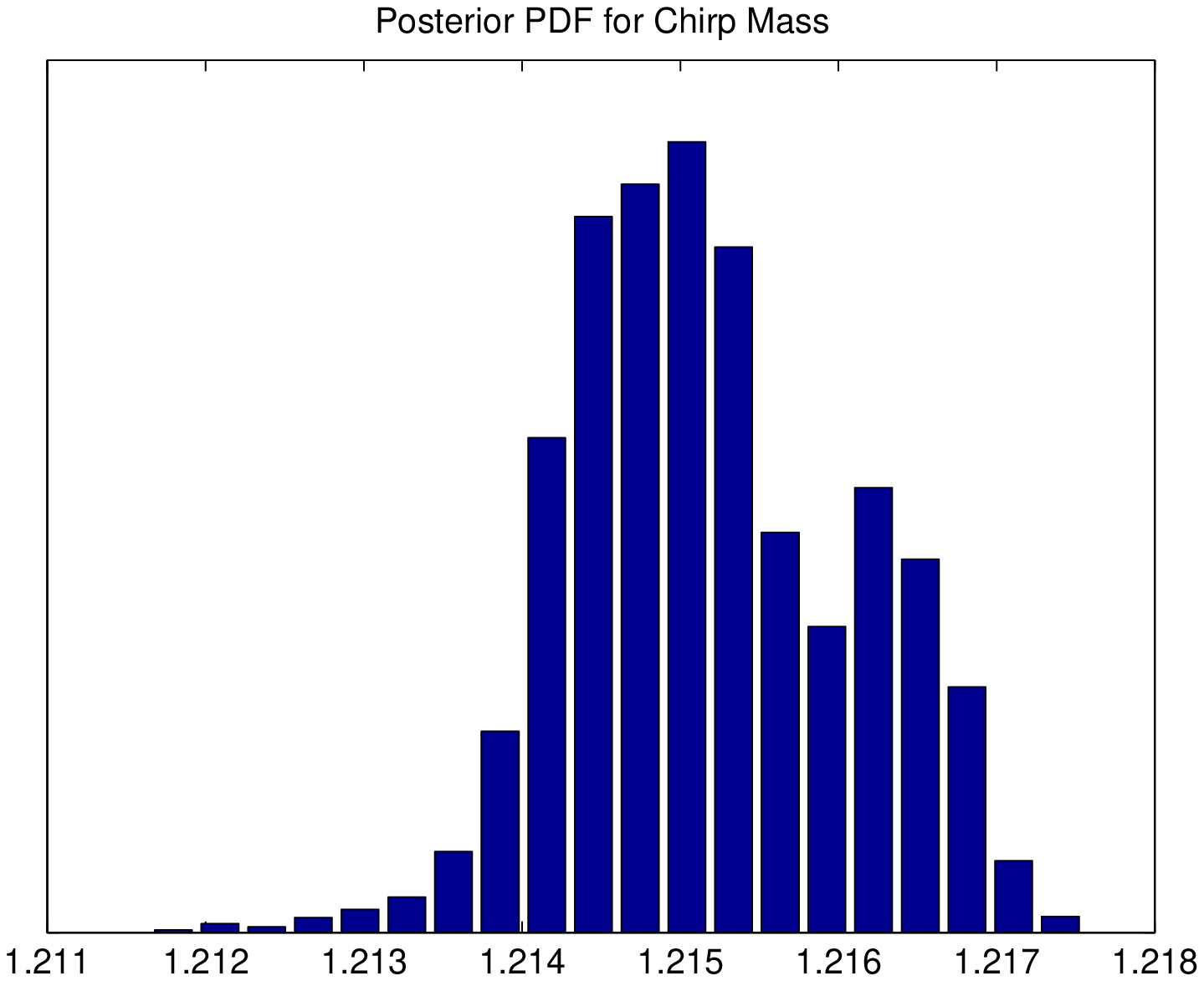}
\hspace*{-0.8cm}
\includegraphics[height=5.cm]{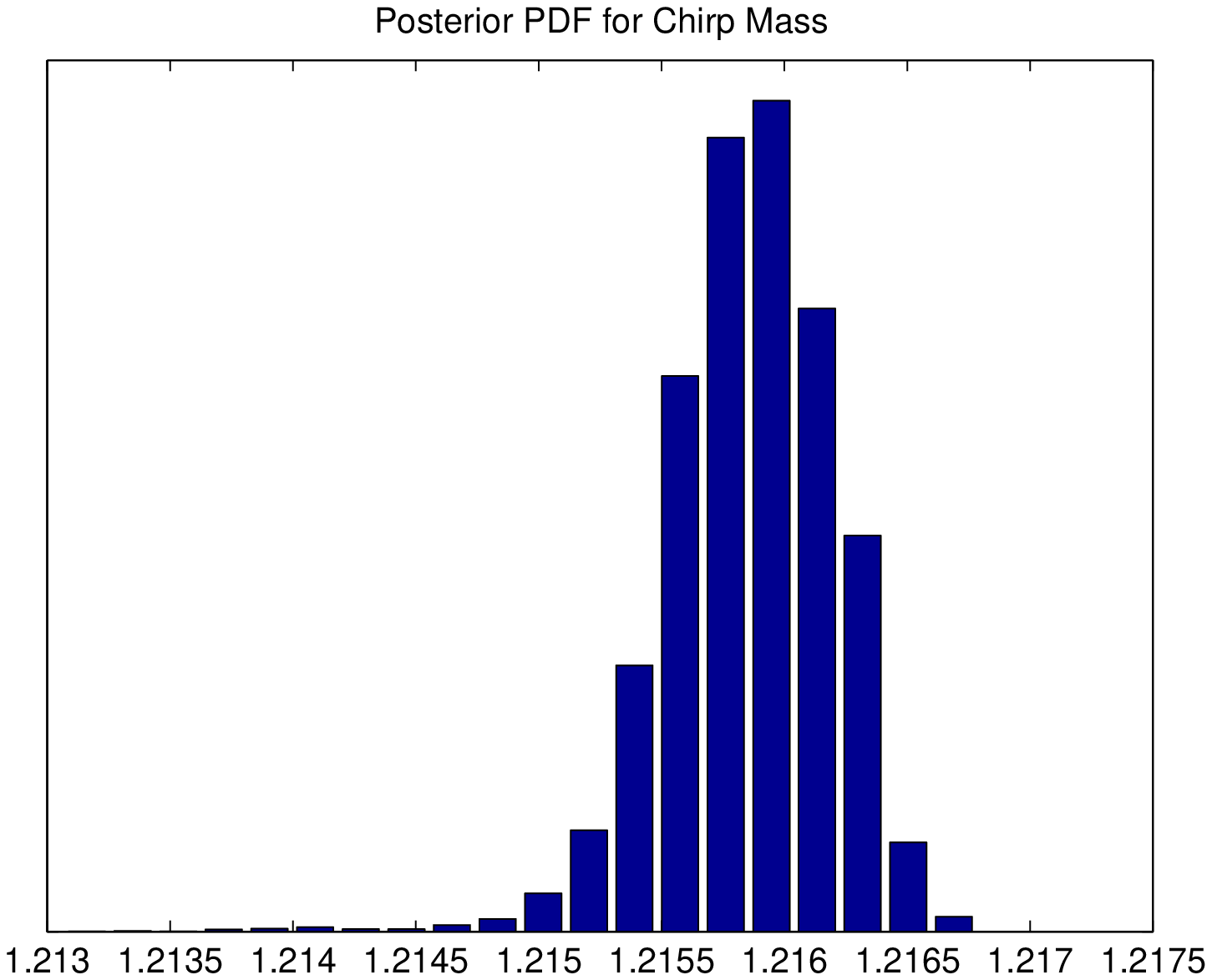}
\vspace*{-.6cm}
\end{center}
\caption{Two examples of posterior PDFs produced by the MCMC for the chirp 
mass. Left: LIGO produced signal at [1~$M_{\odot}$, 2~$M_{\odot}$] (chirp 
mass = 1.2167~$M_{\odot}$), PDF mean= 1.2152, RMS = 0.298. Right: Virgo 
produced signal at [1.4~$M_{\odot}$, 1.4~$M_{\odot}$ (chirp mass 
=1.2188~$M_{\odot}$), PDF mean=1.2158, RMS = 0.185.}
\label{masses_MCMC}
\end{figure}

As a post-processing on selected data sections, a Bayesian parameter 
estimation routine was also applied to the sections 
of data where the inspiral triggers were found. A Markov chain Monte 
Carlo (MCMC) routine using a Metropolis-Hastings algorithm generated 
estimates for the posterior probability distribution functions (PDFs). 
This method, which is designed to find 2.5 post Newtonian inspiral 
signals, is described in~\cite{MCMC1,MCMC2}. Figure~\ref{masses_MCMC} 
shows examples of the MCMC 
produced posterior PDFs for the chirp mass for two of the signals 
analyzed. For the LIGO produced signal with $m_1$=1~$M_{\odot}$, 
$m_2$ =2~$M_{\odot}$ (chirp mass = 1.2167~$M_{\odot}$) the posterior PDF 
for the chirp mass overlaps the injected 
value. For the Virgo produced signal with $m_1$=$m_2$ = 1.4~$M_{\odot}$ 
(chirp mass =1.2188~$M_{\odot}$) the mean of the posterior PDF for the 
chirp mass differs by 
0.23\% from the injected value. This slight difference, also seen with 
the LIGO produced signals with $m_1$=$m_2$=1.4~$M_{\odot}$, 
is likely due to the different 
nature of the signals; the generated signals were 2.0 post-Newtonian in 
the time domain, while the MCMC searches for 2.5 post-Newtonian in the 
frequency domain. The fact that there is not a complete one to one match 
between these two signals is reflected in these estimates.
\section{Conclusion}
In this first project based on simulated data, the LIGO-Virgo joint working 
group has had the opportunity to run analysis pipelines from the two 
collaborations on data sets produced by both sides. In doing so, it has been 
possible to gain better understanding and~-~most importantly~-~confidence 
in each other's detection procedures, since both analysis pipelines have 
shown to detect the same events with comparable parameters. 
This project has thus established the grounds for future work
toward collaborative data analysis in the search for inspiral gravitational 
wave signals.
\section*{Acknowledgements}
LIGO Laboratory and the LIGO Scientific Collaboration gratefully aknowledge 
the support of the United States National Science Foundation for the 
construction and operation of the LIGO Laboratory and for the support of 
this research.

\end{document}